\documentclass[letterpaper, 10 pt, conference]{ieeeconf}  

\IEEEoverridecommandlockouts                              

\usepackage{mathptmx} 
\usepackage{times} 
\usepackage{amssymb}  
\usepackage{graphicx}
\usepackage{amsmath, nccmath}
\usepackage{tabularx,ragged2e,booktabs,caption}
\usepackage{geometry}

\title{\LARGE \bf
	A Testbed for a Smart Building: Design and Implementation
}

\author{Roja Eini$^{1}$, Lauren Linkous$^{2}$, Nasibeh Zohrabi$^{3}$, Sherif Abdelwahed$^{4}$
	\thanks{$^{1}$Roja Eini is PhD student of Department of Electrical Engineering,
		Virginia Commonwealth University, Richmond, VA, USA
		{\tt\small einir@vcu.edu}}%
		\thanks{$^{2}$Lauren Linkous is PhD student of Department of Electrical Engineering,
		Virginia Commonwealth University, Richmond, VA, USA
		{\tt\small linkouslc@vcu.edu}}%
		\thanks{$^{3}$Nasibeh Zohrabi is faculty of Department of Electrical Engineering,
		Virginia Commonwealth University, Richmond, VA, USA
		{\tt\small zohrabin@vcu.edu}}%
	\thanks{$^{4}$Sherif Abdelwahed is faculty of Department of Electrical Engineering,
		Virginia Commonwealth University, Richmond, VA, USA
		{\tt\small sabdelwahed@vcu.edu}}%
}

\begin{document}

\maketitle
\thispagestyle{empty}
\pagestyle{empty}

\begin{abstract}
This paper addresses the design and implementation of a smart building prototype. The implementation utilizes Internet of Things (IoT) solutions to collect, analyze, and manage data from building systems in a smart city environment. The developed smart building prototype is capable of real-time interactions with the residents. The main objective is to adapt the building settings to the residents’ needs and provide the maximum comfort level with minimum operational costs. For this purpose, building parameters are collected via the sensors and transferred to a database in real-time, which can be accessed or visualized upon the users' need. Environment properties such as temperature, light, humidity, audio, video, surveillance, and access status are managed through a model-based controller. The developed testbed and control scheme are generic and modular. The prototype can also be utilized for testing cyber-physical systems' features and challenges. 
\end{abstract}

\section{Introduction}\label{sec:intro}
Smart buildings are essential elements of smart city infrastructure. An intelligent building integrates technologies to create a facility that is safer, more comfortable, efficient, and productive for its occupants. The main aim in a smart building is to provide a convenient environment for the residents while reducing energy consumption and operational costs, using modern design processes and technologies [1-5].   \\ \indent
A smart building consists of sensing and data acquisition, data analysis, data storage, and visualizations. Sensing devices are placed in various positions in the house to gather information related to living conditions (e.g., temperature, humidity, movement, power utilization, and appliances settings). The collected data is transmitted to a local computation node (raspberry pi) where the data is analyzed, and commands/controls are generated. The data and control commands are stored in a central server platform. For visualization, the collected data is displayed through a device such as a smartphone or TV.\\  \indent
Most studies on smart buildings are based on digital simulations or computer-based representations [6], [7]. Moreover, most of the constructed smart buildings are single-floor testbeds with limited components [8], [9]. Digital simulations have the benefit of flexible and quick design, however, they may not provide a real-world description of the system and result in inaccurate outputs. The contribution of this paper is the introduction of a control system paired with a multi-floor, multi-room testbed where the sensors network has not been custom tailored to individual rooms. As such, the developed control system can be modified and used for any cyber-physical or smart city project, and customization of the sensors network is straightforward due to the modular nature of the testbed. Multi-floor, 3D, physical testbed provides a system for both the design and implementation of smart buildings that a computer simulation may lack. \\ \indent
This paper discusses the design and implementation of a smart building. The constructed testbed is a four-story building equipped with sensors and actuators. The aim is to enhance the residents' comfort and save time and energy in a smart framework. Building parameters are measured through a set of embedded sensors. The collected data is processed using a local computation node (processor) on each floor, and the commands are sent to the interfaces. Finally, the system's actuators run based on the processor's commands. The building components' status is accessible at any time. Most of the communications and data transfer is wireless. The experimental results demonstrate the efficiency of the design methodologies for building automation.  \\ \indent
The remainder of the paper is organized as follows: Section~\ref{sec:structure} introduces the smart building structure and its components. Section~\ref{sec: objectives} describes the objectives of the project. The design and implementations of different units in the smart building project are explained in section~\ref{sec: implementation}. Section~\ref{sec:experimental} illustrates the experimental results in detail. Finally, the concluding remarks and future works are presented in Section~\ref{sec:conc}.
\section{System Structure}\label{sec:structure}
\begin{figure}[t]
	\centering 
	\includegraphics[height=5cm, width=6cm]{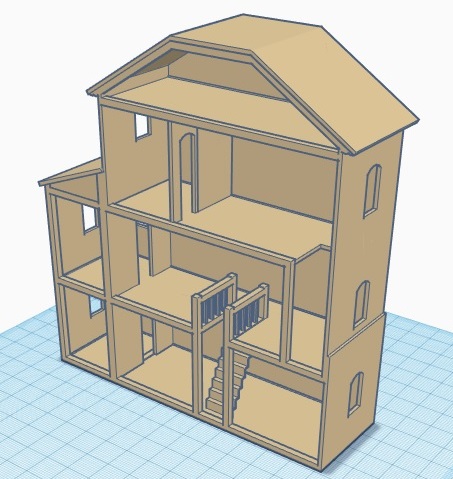}
	\caption{The smart home's CAD plan}
	\label{fig1}
	\vskip 5mm
	\includegraphics[height=5cm, width=7cm]{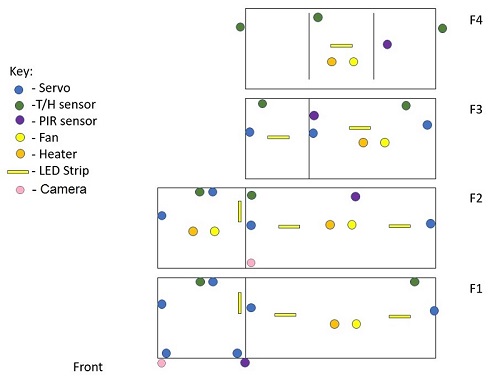}
	\caption{The position of all the sensors and actuators in each floor (top view)}
	\label{fig3}
\end{figure}

The system is a four-story building with differing floor plans. Fig. \ref{fig1} shows the 3D plan of the building prototype. The rooms on each floor are designated as follows: the ground floor contains an open living area divided between a kitchen and a dining room. The stairs and elevator lead to the second floor, where there is a large living area with access to an enclosed sun-room. The third floor is divided between a bedroom and bathroom, with elevator access in the bedroom. Finally, the fourth floor is an isolated attic space. \\ \indent
The frame of the testbed is prefabricated using composite wood sheets and wooden dowels. The room divisions and doors are constructed using 3D printing from Birchwood. The building's elements are built modular such that they can be installed and removed if required.
The list of the sensors and actuators used in the testbed are shown in Table~\ref{table1}. Furthermore, the position of all the actuators and sensors are sketched in Fig.~\ref{fig3}. Table~\ref{table1} does not include some other components used in the testbed, such as Pi ribbon cables, converter cables, jumper wires, IC sockets, proto-boards, capacitors, and resistors. 
Furthermore, 
\begin{table}
	\caption{List of the smart home components}
	\label{table1}
	\begin{tabular}{ccl}
		\toprule
		Component&Part number\\
		\midrule
		Micro Servo Motor&SG92R\\ 
		Cooling/Heating Peltier&TEC1 12706 TEC\\ 
		PC Fan&259-1790-ND\\ 
		Heat sink&AE10830-ND\\ 
		LED strip&NeoPixel\\ 
		Camera&9SIACJ462G8129\\ 
		Temperature/Humidity sensor&DHT22\\ 
		PIR motion sensor&1528-1991-ND\\ 
		Relay&OSA-SH-205DM5,600-ND\\ 
		Wifi module&HUZZAH ESP8266, 2471\\ 
		Raspberry Pi&Pi3 model B+ \\ 
		Pi SD card&128GB\,\,SanDisk\\ 
		AC/DC Converter 5V 60W&1470-1103-ND\\ 
		AC/DC Converter 12V 120W&1470-1088-ND\\ 
		Voltage regulator&SG7805\\ 
		\bottomrule
	\end{tabular}
\end{table}
\section{Project Objectives}\label{sec: objectives}
The main objectives of this project are as follows: 
\subsection{Improving occupant comfort}
The prominent objective of a smart home is to provide the maximum comfort level for the residents [9]. First, a highly secure and reliable surveillance system is necessary for the building. Second, the home parameters (light intensity, temperature, humidity, access status, and video/audio) should adapt to the residents' preferences to provide maximum comfort level with minimum operational costs. Moreover, it is required to eliminate the parameters' oscillations (e.g., the temperature and humidity fluctuations around the desired levels). 
This paper aims to cover the following comfort aspects of a smart home:
\begin{itemize}
	\item Heating/cooling system: \\
	Smart thermostats are designed such that they learn and adjust to the users’ preferences. They learn to function based on the occupancy, time of day, and climatic conditions. The desired temperatures/humidity can be set for each area manually through the dashboard, or automatically through the control algorithm. The smart heating/cooling system adjusts the temperature based on a temperature pattern to help the residents sleep and wake up at specific times.
	\item Lighting: \\
	The lighting system can be properly customized from the dashboard. Indoor and outdoor smart lights turn on when residents enter the building, preventing them from entering a dark house. Outdoor lights paired with PIR motion sensors are used to improve the home security; e.g., by mimicking the routine while the residents are not home to keep the burglars away. The lighting system learns behavior as time goes by so it adjusts to the users' preference.
	\item Communication and connection: \\
	Smart hubs are the smart home brain; they provide an interface for connecting and automating multiple devices and their functionalities. Smart hubs help all the devices in the smart home to speak the same language, making them a great starting point for any smart home remodel. High-speed Internet connection is required to transfer the information to and from the server.     
	\item Security: \\
	A security camera records a video for several seconds while someone is at the front door. Theft and vandalism can be prevented by a smart siren to scare away intruders with a loud sound. Motion sensors paired with video and audio actuators can be used to monitor pets left at home. Smoke/CO2 detectors, door/window sensors, temperature sensors, and vibration sensors can send alerts whenever an unusual change happens. Devices behind a smart firewall, including smartphones, computers, and other smart home products, are shielded from unauthorized access by monitoring the traffic data of connected devices.    
	\item Audio/video: \\
	The smart audio system can be customized to play a white noise in a room when it is time to sleep, or to play a wake up music in the morning. Combined with motion sensors, the audio system can be used to warn the residents when a pet is crossing into an off-limit zone.     
	\item Door/window: \\
	Residents can control doors and windows status remotely. Also, doors/window can be locked or unlocked~automatically. Doors/windows status changes can be sent to the residents to monitor their children. 
\end{itemize}
\subsection{Reducing resource usage}
The smart home project aims to improve the energy consumption by adding features that are typically lacking in a static Building Management System (BMS). Energy usage in a smart building can be reduced using efficient model-based predictive controllers (MPC) [10]. Moreover, a decentralized framework can manage lights, data streaming, temperature, humidity, and other related parameters in a way that it allows for the individual modifications and maintenance in each section [10]. 
\subsection{An open source, generic, and modular scheme}
This project aims to develop a generic and modular scheme for the smart building, such that it can be modified whenever needed. Moreover, the developed smart building model and its control scheme can be utilized as a prototype for further tests in the area of cyber-physical systems' features and challenges (e.g., cyber security). 
\section{Project Implementation}\label{sec: implementation}
\begin{figure}[h]
	\centering
	\includegraphics[height=2.6cm, width=8cm]{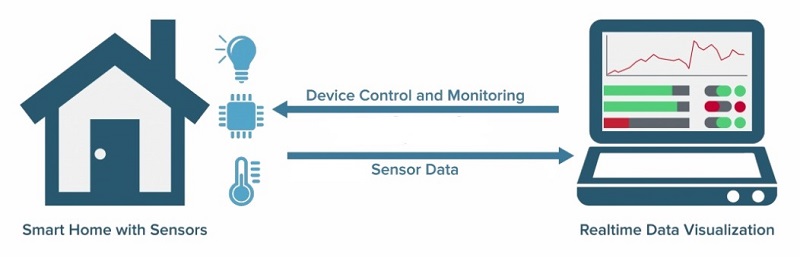}
	\caption{General block diagram of a smart home realization}
	\label{fig40}
\end{figure}

The smart home project realization is categorized into three units: control and data analytics unit, communication and data unit, and visualization unit. Fig. \ref{fig40} shows a general block diagram of a smart building realization.
\subsection{Control and data analytics unit}
\begin{figure}[h]
	\centering
	\includegraphics[height=4cm, width=6cm]{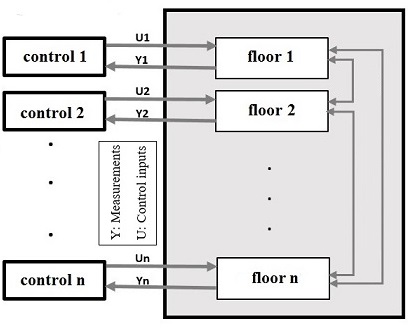}
	\caption{Simple block diagram of a decentralized model-based control}
	\label{fig4}
\end{figure}

The control and analytics unit is developed on a Raspberry Pi module. The essential benefit of using the Raspberry Pi in a home automation system is its open-source operating system, Raspbian, which is based on Debian. Moreover, the power consumption of the Raspi board is very low (between 0.4 to 1.2  W depending on the model) compared to the similar micro-controller boards. \\ \indent
Furthermore, the control framework is developed~modular, which means that all the actuators get commands and settings from the base classes. Therefore, it is easy to adapt the control framework to new system architecture.  \\ \indent 
For controlling the heating/cooling system, a decentralized controller is designed. In a decentralized architecture, one controller is applied on each floor. Hence, each floor (including its components) is considered a subsystem. Thus, the decentralized controller is more reliable because each floor (subsystem) is considered individually for modification or maintenance. A simple block diagram of the decentralized control scheme is shown in Fig.~\ref{fig4} [10]. 

A model-based predictive controller is used as a control strategy. The system's previous input/output information and the environment data are fed into the control unit to generate the control law. For more information on the decentralized predictive control algorithm for building automation see [10]. 

Hence, the light and heating/cooling system performance using the predictive framework is recognizably faster. Utilizing a model-based decentralized strategy improves the control performance in tracking the desired building settings and maintaining the residents' comfort. Therefore, energy consumption and expenses would be decreased significantly using the decentralized model predictive scheme.
\subsection{Communication and data storage unit}
Interoperability is considered as an important issue in the communication between system components. It is more efficient for a device or a system to communicate with other components using open protocols. In this way, the device will not be restricted to vendor-specific solutions and their subsequent non-competitive pricing, and therefore, a change or addition can be made easily in the future. So, the communication unit of this smart home project is developed open source since this makes the modifications easier. \\ \indent
In the communication unit, the Wifi modules are connected to the Raspberry Pi wireless. A basic network between the Raspberry Pi and components is implemented using the dynamic host configuration protocol. In this stage of the project, the Raspberry Pi is used as the computation node where several nodes (i.e., addressable devices) are connected to it. \\ \indent
Therefore, the Pi itself is configured as an access point, so that the sensors and actuators can communicate directly with it. The components communicate through message queuing telemetry transport (MQTT) protocol. Each device in the network has an embedded ESP8266 Wifi Module. All the components send data to the Pi, which is then saved in a local database before being transferred to an off-line server. The database consistently stores the sensor measurements, actuators' status, control commands, analysis, and visualization unit's data. 
\subsection{Visualization unit}
The Dashboard is designed to display analytical data such as light intensity, temperature, humidity, surveillance, door status, and camera steaming. The Dashboard is designed as a web server that pulls data from a database, processes it for display, and displays it in real-time over the IoTNet on a central Dashboard television, as well as the end user's personal device. The Dashboard's display elements are automatically formatted to the viewing device (e.g., a TV, a laptop, or a smart-phone). The display elements can be added or removed from the user's display via a drop-down select menu, allowing for unlimited flexibility for debugging and analyzing data from different projects in real-time.  \\ \indent
The Dashboard application is designed on the IoTNet central server, in an Ubuntu virtual machine (VM). The Dashboard application runs on a shared-kernel imaging program called Docker. Docker images are more flexible and easier to modify than traditional virtual machines. In the VM, a shared-kernel image hosted by Docker contains a stripped-down Linux server with Python 3.6, SQLite, and Apache for web hosting. The Dashboard exists in three parts within the Docker container: A Data Parser, written in Python, that connects to the IoTNet Server's database controller, a local SQLite Database that stores the parsed data, and a Web App, hosted on Apache, that generates HTML in JSON form for display and user interfacing.  \\  \indent
 \noindent
\begin{figure}[h]
	\begin{center}
		\parbox{3in}{
			\centering
			\includegraphics[height=4cm, width=6.8cm]{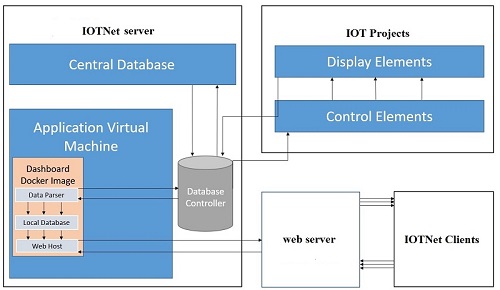}
		}
		\caption{Functional diagram of the implemented visualization unit}
		\label{fig50}
	\end{center}
\end{figure}  

The Dashboard runs on an Ubuntu 18.04 image that contains Dash, Plotly, Flask, Pandas, Apache, and SQLite packages. The WebSocket protocol is used as a way to stream data in real time from the server to the users on the network. Furthermore, the data on the Dashboard is accessible whenever the user requires. Fig. \ref{fig50} shows the block diagram of the designed visualization panel for the smart building testbed.
\section{Experimental Results}\label{sec:experimental}
\begin{figure}[thpb]
	\begin{center}
		\parbox{3in}{
			\centering
			\includegraphics[height=5cm, width=7cm]{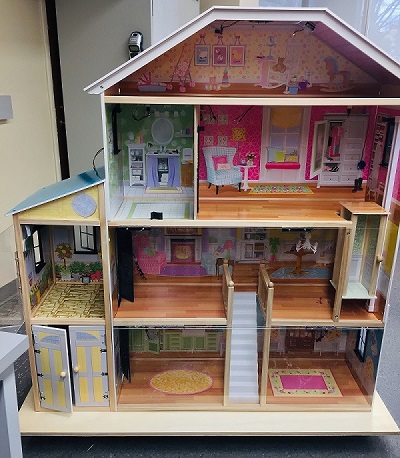}
		}
		\caption{Picture of the smart building testbed}
		\label{fig100}
	\end{center}
	\begin{center}
		\parbox{3in}{
			\centering
			\includegraphics[height=4cm, width=7cm]{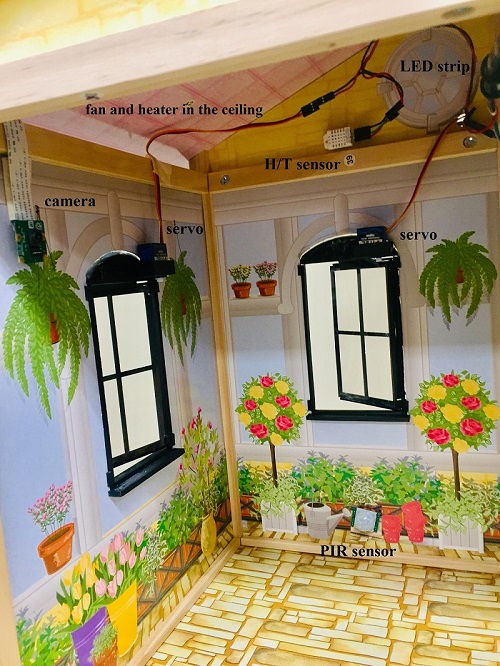}
		}
		\caption{Picture of one room including its actuators and sensors}
		\label{fig101}
	\end{center}
	\begin{center}
		\parbox{3in}{
			\centering
			\includegraphics[height=4cm, width=7cm]{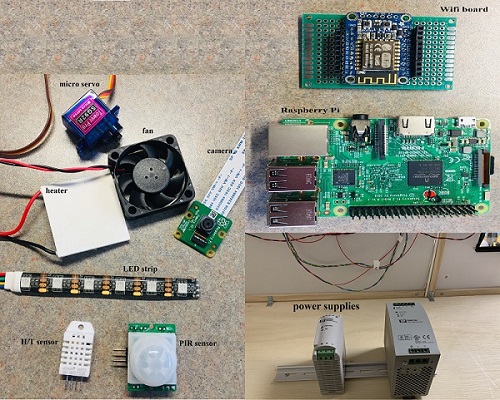}
		}
		\caption{Picture of sensors, actuators, sources, and control boards used in the smart building}
		\label{fig102}
	\end{center}
\end{figure}

This section provides the experimental results of the implemented smart home. Fig.~\ref{fig100} shows the picture of the constructed building. According to Fig. \ref{fig100}, each pane is hinged so that floors can be accessed individually without disturbing the conditions in the rest of the building. \\ \indent
As can be seen in Fig. \ref{fig101}, the sensors and actuators are mounted on the walls, doors, windows, elevator, and ceilings. Each room of the building, excluding the garage, has at least one strip of LEDs as a light source, a DTH22 as a temperature/humidity sensor, a Peltier tile as a heat source, a fan for heat dissipation, PIR sensors to monitor the movements, and micro servos to manage doors/windows status. The garage does not have any temperature control. The front door is controlled by a camera and a larger servo (to compensate for the added weight). Fig. \ref{fig101} shows the larger picture of some actuators and sensors (e.g., PIR sensor, DHT22 sensor, servo motor, fan, heater, LEDs, and camera) mounted in a room.

Sensor readings are measured and sent to a central server by Adafruit HUZZAH ESP8266 breakout boards. Some actuators, such as the micro servos, are also connected to the ESP boards. LEDs, relays, cameras, and fans are connected directly to the computation node (Raspberry Pi). The camera records videos/capture pictures when the near-by PIR sensor detects movement. The camera can also stream videos and pictures upon the resident's command. The controller unit (Raspberry Pi module), interface modules (e.g., the voltage regulator modules, power isolation modules, and Wifi modules), and the power supplies are mounted at the back of the building. Fig. \ref{fig102} shows the actuators, sensors, voltage sources, and control boards used in the testbed. \\ \indent
The circuits for the 12V and 5V components are kept isolated by relays, and as a precaution in the event of a high current, all micro-controllers are isolated from each other with fuses. Power can be supplied to any combination of individual floors. All components are in parallel and connected via a modular wiring harness system for easy adaptation. \\ \indent
For the lighting system, each LED strip contains 5 LEDs. Different voltage levels are applied to generate different colors for each of the living areas. The voltage range required is between 1.4 to 5 V, and the maximum current needed per LED strip is 0.29 A. 

\begin{figure}[thp]
	\begin{center}
		\framebox{\parbox{3in}{
				\centering
				\includegraphics[height=5cm, width=7cm]{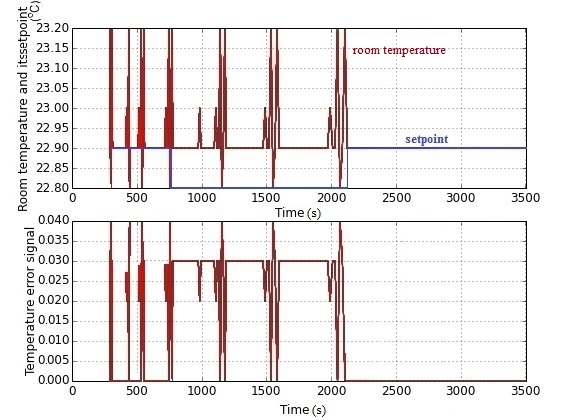}
		}}
		\caption{First-floor temperature trajectory}
		\label{fig20}
	\end{center}
	\begin{center}
		\framebox{\parbox{3in}{
				\centering
				\includegraphics[height=5cm, width=7cm]{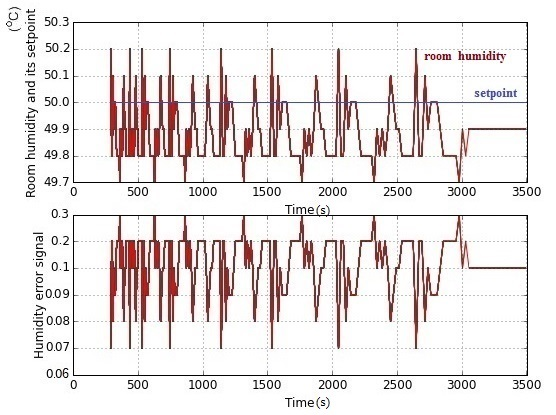}
		}}
		\caption{First-floor humidity trajectory}
		\label{fig21}
	\end{center}
	\begin{center}
		\framebox{\parbox{3in}{
				\centering
				\includegraphics[height=5cm, width=7cm]{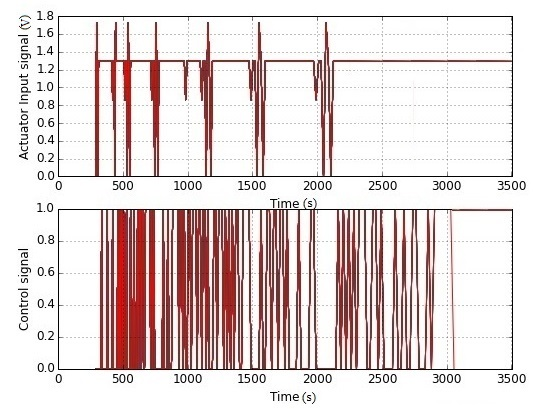}
		}}
		\caption{First-floor actuator input and control signal}
		\label{fig22}
	\end{center}
\end{figure}

The temperature and humidity in each floor are regulated based on the residents' desired levels. A decentralized model-based predictive control is developed in Python 3.6 and loaded on the micro-controller through Raspbian. The modules that are most used are numpy, csv, pylab, matplotlib, time, RPi.GPIO, string, and scipy. The sensors' data is updated every 5 seconds, and the control inputs are generated at the same rate. Fig. \ref{fig20} shows the trajectories of the first-floor temperature versus the desired temperature. Fig. \ref{fig21} shows the humidity trajectory and the desired humidity signal. The average temperature error and average humidity error are 2.5\% and 10\%, respectively. Fig. \ref{fig22} illustrates the control law and the input signal generated for the actuators (fan and heater). 
\section{Conclusions and Future Works}\label{sec:conc}
The integration is often overlooked during the design phase of a smart building. In this study, the smart home is implemented modular and generic, such that the integration is actualized. Moreover, the decentralized control scheme is developed in the building to ease system modification and maintenance. The building's features (surveillance, humidity/temperature level, light intensity, and data streaming) are managed in real-time, through the model-based predictive controller. The residents' comfort, which is the prominent aim of a smart home, is maintained using the developed structure.

The testbed opens up many opportunities for future expansions for the practical research of smart city and cyber-physical systems. Smart appliances, such as smart energy saver appliances, and adjustable furniture (standing/sitting tables), smart plugs, automated garbage classification, and home entertainment can be added to the project in the future. Furthermore, the controller framework and the system can be developed for larger buildings.

\end{document}